\documentclass[journal]{IEEEtran}
\usepackage{calc}

\usepackage{multirow}
\usepackage{cite}
\usepackage{graphicx,subfigure}
\usepackage{psfrag}
\usepackage{amsmath,amssymb}
\usepackage{color}
\usepackage{tikz}
\usepackage{pgfplots}
\usepackage{mathtools}

\usetikzlibrary{snakes}

\newenvironment{proof}{\begin{IEEEproof}}{\end{IEEEproof}}

\newtheorem{theorem}{Theorem}
\newtheorem{corollary}{Corollary}[theorem]

\newtheorem{lemma}{Lemma}

\newcommand{\bit}{\begin{itemize}}
\newcommand{\eit}{\end{itemize}}

\newcommand{\bc}{\begin{center}}
\newcommand{\ec}{\end{center}}

\newcommand{\ba}{\begin{array}}
\newcommand{\ea}{\end{array}}

\newcommand{\beq}{\begin{equation}}
\newcommand{\eeq}{\end{equation}}

\newcommand{\beqn}{\begin{equation*}}
\newcommand{\eeqn}{\end{equation*}}

\newcommand{\bean}{\begin{eqnarray*}}
\newcommand{\eean}{\end{eqnarray*}}
\newcommand{\bea}{\begin{eqnarray}}
\newcommand{\eea}{\end{eqnarray}}

\def\Z{\mathbb{Z}}

\def\E{\mathbb{E}}

\def\hv{\boldsymbol{h}}

\def\xv{\boldsymbol{x}}

\begin{document}
\sloppy

\title{The Synergistic Gains of Coded Caching \\ and Delayed Feedback}
\author{Jingjing Zhang and Petros Elia
\thanks{The authors are with the Mobile Communications Department at EURECOM, Sophia Antipolis, 06410, France (email: jingjing.zhang@eurecom.fr, elia@eurecom.fr).
The work of Petros Elia was supported by the the ANR Jeunes Chercheurs project ECOLOGICAL-BITS-AND-FLOPS.}
\thanks{An initial version of this paper has been reported as Research Report No. RR-15-307 at EURECOM, August 25, 2015, http://www.eurecom.fr/publication/4723, and can also be found in \cite{ZEarxiv:15}.}
}


\maketitle

\begin{abstract}
In this paper, we consider the $K$-user cache-aided wireless MISO broadcast channel (BC) with random fading and delayed CSIT, and identify the optimal cache-aided degrees-of-freedom (DoF) performance within a factor of 4. The achieved performance is due to a scheme that combines basic coded-caching with MAT-type schemes, and 
which efficiently exploits the prospective-hindsight similarities between these two methods. This delivers a powerful synergy between coded caching and delayed feedback, in the sense that the total synergistic DoF-gain can be much larger than the sum of the individual gains from delayed CSIT and from coded caching.

The derived performance interestingly reveals --- for the first time --- substantial DoF gains from coded caching, even when the (normalized) cache size $\gamma$ (fraction of the library stored at each receiving device) is very small. Specifically, a microscopic $\gamma \approx e^{-G}$ can come within a factor of $G$ from the interference-free optimal. For example, storing at each device only a \emph{thousandth} of what is deemed as `popular' content ($\gamma\approx 10^{-3}$), we approach the interference-free optimal within a factor of $ln(10^3) \approx 7$ (per user DoF of $1/7$), for any number of users. This result carries an additional practical ramification as it reveals how to use coded caching 
to essentially buffer CSI, thus partially ameliorating the burden of having to acquire real-time CSIT.

\end{abstract}

\section{Introduction}

In the setting of broadcast-type communication networks where one transmitter serves the interfering requests of more than one receiving user, recent work in \cite{MN14} showed how properly-encoded caching of content at the receivers, and proper encoding across different users' requested data, can provide increased effective throughput and a reduced network load. This was achieved by creating --- through coding --- multicast opportunities where common symbols are simultaneously needed by more than one user, even if such users requested different data content.
This \emph{coded caching} approach has since motivated different works in \cite{WLTL:15,MND13,JTLC:14,MN:15isit,TW:15,GKY:15,JTLC:15,SJTLD:15,ZFE:15,SMK:15,HA:2015,WLG:15,APPV:15,HuangWDY015,BBD:15,MCOFBJ:14,HKD:14,HKS:15,DBAD:15} which considered the utilization of coded caching over a variety of different settings, including the recent concurrent works in~\cite{GKY:15} that considered the cache-enabled broadcast packet erasure channel with ACK/NACK feedback, and the preliminary work in \cite{ZFE:15} that considered caching with imperfect-quality feedback.

Part of our motivation here is to explore the connection between coded caching and communications with imperfect feedback. Intuitively both cases face parallel problems: a transmitter with complete data knowledge, must retroactively compensate for only having partial knowledge of the `destination', may this be the identity of the receiving user the `next day', or the partially known channel. These connections will eventually allow for a non-separability property which we here translate into substantial synergistic gains between delayed feedback and coded caching. These gains are pertinent because there is a real need to boost the performance effect of generally modest cache sizes, and because delayed CSIT is often the only feedback resource that is available in larger networks with rapidly fluctuating channel states.

\subsection{Caching-aided broadcast channel model}

We consider the setting of the symmetric multiple-input single-output (MISO) BC where a transmitter that is equipped with $K$ antennas, communicates to $K$ single-antenna users. The transmitter has access to a library of $N$ distinct files $W_1, W_2, \cdots, W_N$, each of size $|W_n| = f$ bits. Each user $k \in {1,2,...,K}$ has a cache $Z_k$ with size $|Z_k| = Mf$ (bits), and this size takes the normalized form \[\gamma := \frac{M}{N}.\]
The communication has two phases, the placement phase and the delivery phase. During the first phase (off-peak hours), the caches $\{Z_k\}_{k=1}^{K}$ at the users are pre-filled with the information from the $N$ files $\{W_n\}_{n=1}^{N}$. During the second phase, the transmission commences when each user $k$ requests a single file $W_{R_k}$ out of the library.

In this setting, the received signals at each user $k$ take the form
\[y_{k}=\hv_{k}^{T} \xv + z_{k}, ~~ k = 1, \dots, K\]
where $\xv\in\mathbb{C}^{K \times 1}$ denotes the transmitted vector satisfying a power constraint $\E(||\xv||^2)\leq P$, where $\hv_{k}\in\mathbb{C}^{K\times 1}$ denotes the vector fading coefficients of the channel of user $k$ and where $z_{k}$ represents unit-power AWGN noise at receiver $k$.
\begin{figure}[t!]
  \centering
\includegraphics[width=0.8\columnwidth]{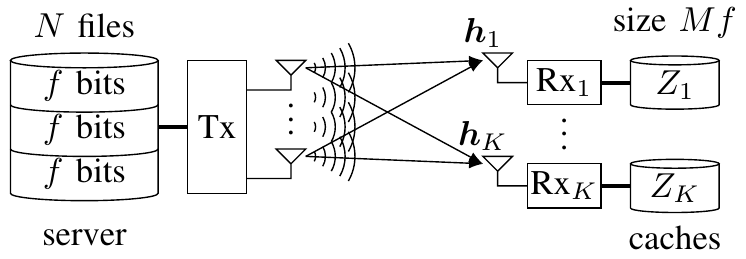}
\caption{Cache-aided $K$-user MISO BC.}
\label{fig:model}
\end{figure}

At the end of the communication, each receiving user $k$ combines the received signals $y_{k}$ --- accumulated during the delivery phase --- with the information available in their respective cache $Z_k$, to reconstruct their desired file $W_{R_k}$.

%

\subsection{Measures of performance}

As in \cite{MN14}, the measure of performance here is the duration $T$ --- in time slots, per file served per user --- needed to complete the delivery process, \emph{for any request}. The wireless link capabilities, and the time scale, are normalized such that one time slot corresponds to the optimal amount of time it would take to communicate a single file to a single receiver, had there been no caching and no interference.
As a result, in the high $P$ setting of interest --- where the capacity of a single-user MISO channel scales as $\log_2(P)$ --- we proceed to set $f = \log_2(P) $
which guarantees that the two measures of performance, here and in \cite{MN14}, are the same and can thus be directly compared\footnote{
We note that setting $f = \log_2(P)$ is simply a normalization of choice, and does not carry a `forced' relationship between SNR and file sizes. The essence of the derived results would remain the same for any other non-trivial normalization.}. Our objective here is to identify caching and transmission schemes that jointly reduce $T$.
A simple inversion leads to the equivalent measure of the per-user DoF
\[d(\gamma) = \frac{1-\gamma}{T} \in [0,1]\]
which captures the joint effect of coded caching and of any feedback resources that might be available\footnote{The DoF measure is designed to exclude the benefits of having some content already available at the receivers (local caching gain), and thus to limit the DoF between 0, and the interference free optimal DoF of 1.}.

\subsection{Notation and assumptions}
We will use $\Gamma := \frac{KM}{N} = K\gamma$ to mean that the sum of the sizes of the caches across all users, is $\Gamma$ times the volume of the $N$-file library. As in~\cite{MN14}, we will consider the case where $\Gamma = \{1,2,\cdots K\}$. We will also use the notation $H_n := \sum_{i=1}^{n} \frac{1}{i}$ to represent the $n_{th}$ harmonic number, and we will use $\epsilon_n := H_n-\log (n)$ to represent its logarithmic approximation error, for some integer $n$. We remind the reader that $\epsilon_n$ decreases with $n$, and that $\epsilon_\infty :=\lim \limits_{n \rightarrow \infty} (H_n -  \log (n)) \approx 0.5772$.
$\binom{n}{k}$ will be the $n$-choose-$k$ operator, and $\oplus$ will be the bitwise XOR operation. We will use $[K]:= \{1,2,\cdots,K\}$. If $\psi$ is a set, then $|\psi|$ will denote its cardinality. For sets $A$ and $B$, then $A \backslash B$ denotes the difference set.
Complex vectors will be denoted by lower-case bold font. We will use $||\xv||^2$ to denote the magnitude of a vector $\xv$ of complex numbers. For a transmitted vector $\xv$, we will use $\text{dur}(\xv)$ to denote the transmission duration of that vector, e.g., $\text{dur}(\xv) = \frac{1}{10}T$ would mean that the transmission of vector $\xv$ lasts one tenth of $T$.
Logarithms are of base~$e$, while $\log_2(\cdot)$ will represent a logarithm of base~2.

\subsubsection{Main assumptions}
Throughout this work, we assume availability of delayed CSIT (as in for example \cite{MAT:11c}, as well as a variety of subsequent works~\cite{YKGY:12d,CE:13it,GJ:12o,CE:12d,KYG:13,CYE:13isit,VV:09,TJSP:12,LH:12,HC:13}) \nocite{KPR:99,BGW:10} where now the delayed estimates of any channel, can be received at the transmitter, without error but with arbitrary delay, even if this delay renders this CSIT completely obsolete. We hasten to note that delayed CSIT here is not meant to offer an additional performance boost by itself, but rather is employed solely as a tool that will link coded-caching to communications with non-perfect feedback.

We will also ask that each receiver knows their own channel perfectly. We further adhere to the common convention (see for example~\cite{MAT:11c}) of assuming perfect and global knowledge of delayed channel state information at the receivers (delayed global CSIR), where each receiver must know (with delay) the CSIR of (some of the) other receivers. We will assume that the entries of \emph{each specific} estimation error vector are i.i.d. Gaussian.
For the outer (lower) bound to hold, we will make the common assumption that the current channel state must be independent of the previous channel-estimates and estimation errors, \emph{conditioned on the current estimate} (there is no need for the channel to be i.i.d. in time). We will make the assumption that the channel is drawn from a continuous ergodic distribution such that all the channel matrices and all their sub-matrices are full rank almost surely.

\section{Performance of the cache-aided MISO BC}
We begin with an outer (lower) bound on the optimal duration $T^*$. The proof is found in the Appendix.
\vspace{3pt}
\begin{lemma}\label{lem:outer}
The optimal $T^*$ for the $(K,M,N)$ cache-aided $K$-user MISO BC with delayed CSIT, is lower bounded as
\begin{align}
T^* &\geq \mathop {\text{max}}\limits_{s\in \{1, \dots, \text{min} \{\lfloor \frac{N}{M} \rfloor, K\}\}} H_s - \frac{sM}{\lfloor \frac{N}{s} \rfloor}.
\label{eq:outer}
\end{align}
\end{lemma}
\vspace{3pt}



We now proceed with the main result.
\vspace{3pt}
\begin{theorem} \label{thm:MATandMN}
In the $(K,M,N)$ cache-aided MISO BC with $K\leq N$ users, and with $\Gamma \in \{1, 2, \cdots , K-1\}$, then
\begin{align}\label{eq:gammabigBest}
T = H_K-H_\Gamma
\end{align} is achievable and has a gap-to-optimal
\begin{align}
\frac{T}{T^*} < 4
\end{align}
that is less than 4, for all $K$.
\end{theorem}
\vspace{3pt}

\vspace{3pt}
\begin{proof}
The scheme that achieves the above performance is presented in Section~\ref{sec:schemeMATMN}, while the corresponding gap to the optimal performance is bounded in Section~\ref{sec:gapCalculation}.
\end{proof}
\vspace{3pt}

The following corollary offers some insight by adopting the logarithmic approximation $H_n\approx\log (n)$ (which becomes tight as $K$ increases)\footnote{To avoid confusion, we clarify that the main theorem is simply a DoF-type result, that nothing but SNR scales to infinity, and the derived DoF holds for all $K$. The corollaries are simply the approximation of the above expression, under the logarithmic approximation, which becomes tight as $K$ increases.  The corollary is derived directly by approximating the expression in Theorem~\ref{thm:MATandMN} (eq. ~\ref{eq:gammabigBest}) for larger values of $K$.}.
\vspace{3pt}
\begin{corollary}
Under the logarithmic approximation, the above $T$ takes the form
\[ T = \log(\frac{1}{\gamma})\]
and the corresponding per-user DoF takes the form
\begin{align}
d(\gamma) = \frac{1-\gamma}{\log(\frac{1}{\gamma})}.
\end{align}
\end{corollary}
\vspace{3pt}

\subsection{Synergistic DoF gains}
We proceed to derive some insight from the above, and for this we look to the large $K$ regime, where there is no ambiguity on which gains can be attributed solely to coded caching 
(in addition to possible DoF gains due to other resources such as feedback). 
In this regime, what the above says is that the gain that is directly attributed to caching  
\[d(\gamma) - d^*(\gamma=0) \rightarrow \frac{1-\gamma}{\log(\frac{1}{\gamma})} > \gamma, \ \forall \gamma\in(0,1]\]
can substantially exceed\footnote{In this larger $K$ setting, we have
$d_{\textsc{ss}}(\gamma) + d_{\textsc{mat}} \rightarrow \gamma$. 
We clarify that this step is simply the result of a large-$K$ approximation of the corresponding expression from the main theorem. In that sense, $K$ scales after SNR does.
We also recall from~\cite{MAT:11c} that $d^*(\gamma = 0) = \frac{1}{H_K}$ which decreases with $K$.} the typical coded-caching (per-user DoF) gain $\gamma$.

What we also see, again for larger $K$, is that while the individual component settings/algorithms (MAT from \cite{MAT:11c}, and the Maddah-Ali and Niesen (MN) algorithm from~\cite{MN14}) respectively provided individual DoF gains of the form 
$d_{\textsc{mat}} =d^*(\gamma = 0) = \frac{1}{H_K}$ and $d_{\textsc{ss}}(\gamma) = \frac{1-\gamma}{\frac{K(1-\gamma)}{1+K\gamma}} = \gamma+\frac{1}{K}$ (cf.~\cite{MN14}), 
the combination of these two components results in a synergistic 
\[ d(\gamma)  > d_{\textsc{ss}}(\gamma) + d_{\textsc{mat}}, \ \ \forall \gamma\in[0,1]\]
that --- for larger $K$ --- exceeds the sum of the two individual components. This is the first time that such synergistic gains have been recorded. 
The gains become very striking for smaller values of $\gamma$ in which case\footnote{These unbounded gains (as $\gamma$ decreases), also guarantee that the same conclusion holds for any substitute of the algorithm in \cite{MN14}, simply because this
algorithm has a gap to optimal that is bounded (gap less than 12, cf.~\cite{MN14}).} we have that $\frac{1-\gamma}{\log(\frac{1}{\gamma})} \gg \gamma . $ 

These gains in fact imply an exponential (rather than linear) effect of coded caching, in the sense that now a microscopic $\gamma = e^{-G}$ can offer a very satisfactory
\begin{align}
d(\gamma = e^{-G}) \approx \frac{1}{G}
\end{align}
which is only a factor $G$ from the interference-free (cache-free) optimal $d=1$.

\begin{figure}[t!]
  \centering
\includegraphics[width=0.85\columnwidth]{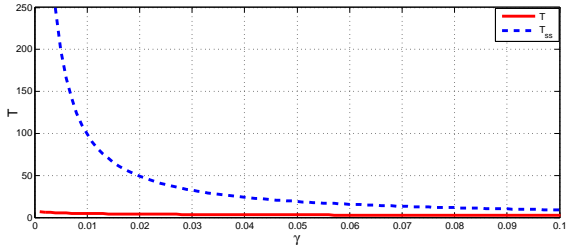}
\caption{Single stream $T_{\textsc{ss}}$ (no delayed CSIT, dotted line) vs. $T$ after the introduction of delayed CSIT. Plot holds even for very large $K$, and the main gains appear for smaller values of $\gamma$.}
\label{fig:comparisonTmn}
\end{figure}



\subsection{Practical ramification: using coded caching to `buffer' CSI}
In addition to the substantial DoF gains that one can get by exploiting synergy, we also note that exploiting this interplay between caching and feedback timeliness, can additionally help alleviate the laborious task of sending feedback under the coherence period constraint. By using a modest $\gamma$, we are essentially endowing the system (for this specific setting that we are considering here) with a seemingly paradoxical ability of online buffering of CSI.
To see this better, consider
\begin{align}\label{eq:gammaIF}
\gamma^{'}_{G}:= \arg\min_{\gamma^{'}}\{\gamma^{'}:  d(\gamma^{'}) \geq \frac{1}{G}\}
\end{align}
which describes the minimum $\gamma$ needed to achieve --- in conjunction with delayed CSIT --- a certain gap $G\geq 1$ from the interference-free (cache-free) optimal (associated to perfect real-time CSIT), for which we can quickly calculate that $\gamma^{'}_{G} = e^{-(G-\epsilon_K+\epsilon_\infty)}$, which for larger $K$, converges to\footnote{The above holds because $T = H_K-H_{\Gamma} = G = \log(\frac{1}{\gamma}) + \epsilon_K-\epsilon_{K\gamma} \leq \log(\frac{1}{\gamma}) + \epsilon_K-\epsilon_\infty$.} the aforementioned $
\gamma^{'}_{G}= e^{-G}.$



\begin{figure}[t!]
  \centering
\includegraphics[width=0.8\columnwidth]{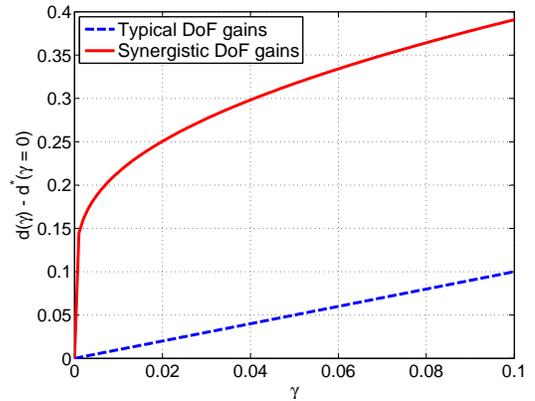}
\caption{Typical gain $d(\gamma)-d^*(\gamma=0)$ attributed solely to coded caching (dotted line) vs. synergistic gains derived here. Plot holds for large $K$, and the main gains appear for smaller values of $\gamma$.}
\label{fig:comparisonTmn}
\end{figure}

\section{Cache-aided prospective-hindsight scheme \label{sec:schemeMATMN}}  

We proceed to describe some of the details of the scheme, and how it combines the coded caching algorithm in \cite{MN14} (placement, folding-and-delivery, and decoding) with the MAT algorithm in~\cite{MAT:11c}. 

\begin{figure}[h!]
  \centering
\includegraphics[width=0.99\columnwidth]{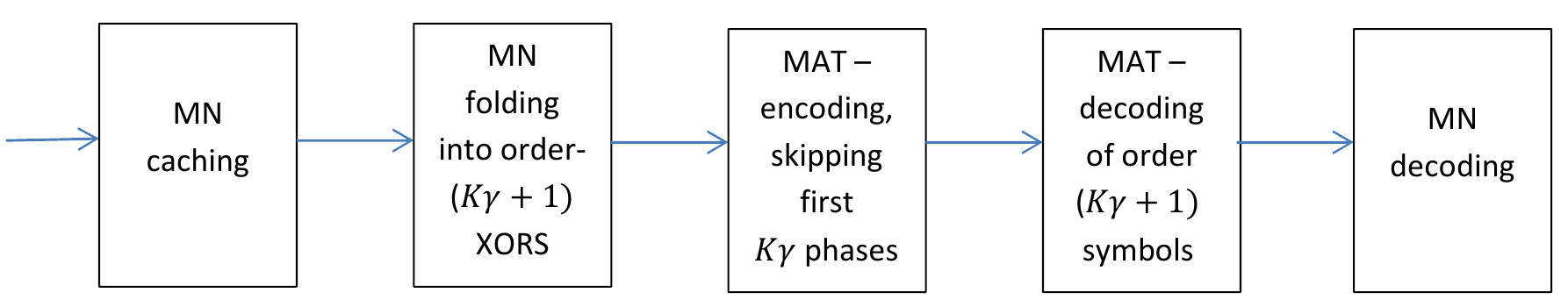}
\caption{Basic composition of scheme. `MAT encoding/decoding' corresponds to the scheme in \cite{MAT:11c}, while `MN caching/folding' corresponds to the scheme in \cite{MN14}.}
\label{fig:simpleScheme}
\end{figure}

\paragraph{Key idea behind the scheme}
First let us briefly describe the idea behind our simple scheme. 
As Figure~\ref{fig:simpleScheme} implies, the scheme starts by first applying the Maddah-Ali and Niesen (MN) sub-packetization based scheme \cite{MN14} in order to place contents (sub-packets) in the caches, and to generate order-$(K\gamma+1)$ messages in the form of XORs of the sub-packets, where each of these XORs is meant for $K\gamma+1$ users. 
These XORs are delivered by the well known MAT method~\cite{MAT:11c}, and in particular the MAT variant that delivers order-$(K\gamma+1)$ messages. 
This allows us to skip the first $K\gamma$ phases of the MAT scheme, which happen to have the longest time duration. 
This is why the impact of even small caches (small $\gamma$) is substantial. 
Upon MAT decoding, we simply proceed with decoding based on the algorithm in~\cite{MN14}. 
In the end, the key idea is that the caching algorithm creates a multi-destination delivery problem that is the same as that which is efficiently solved by the last stages of the MAT scheme.

\subsection{Placement phase}
The placement phase is taken from~\cite{MN14}, where each of the $N$ files $\{W_n\}_{n=1}^{N}$ ($|W_n| = f$ bits) in the library, is equally split into
$\binom{K}{\Gamma}$ subfiles as follows, $W_n = \{W_{n,\tau}\}_{\tau \in \Psi_{\Gamma}}$, where $\Psi_{\Gamma}:= \{\tau \subset [K] \ : \  |\tau| = \Gamma\}$, so each subfile has size
\begin{align}
|W_{n,\tau}| = \frac{f}{\binom{K}{\Gamma}} \ \text{bits}.  \label{eq:WnTauSize} \end{align}

Based on the above, the caches are filled as follows $Z_k=\{W_{n,\tau}\}_{n \in [N], \tau\in \Psi_{\Gamma},k \in \tau}$, 
so that each subfile $W_{n, \tau}$ is stored in $Z_k$ as long as $k\in\tau$.

\subsection{Delivery}
At the beginning of the delivery phase, the transmitter must deliver each requested file $W_{R_k}$, by delivering the constituent subfiles $\{W_{R_k,\tau}\}_{k \notin \tau}$ to the corresponding user $k$. Thus, as in~\cite{MN14}, we must send the entire set $\mathcal{X}_\Psi := \{ X_{\psi}\}_{\psi \in  \Psi_{\Gamma+1}}$, where  each XOR is of the form $X_{\psi} := \oplus_{k \in \psi} W_{R_k,\psi \backslash \{k\}}, \psi \in  \Psi_{\Gamma+1}$. There are $|\mathcal{X}_\Psi|=\binom{K}{\Gamma+1}$
folded messages (XORs), and each has size (cf. \eqref{eq:WnTauSize})
\begin{align} \label{eq:XpsiSize}
|X_{\psi}| & = |W_{R_k,\tau}| = \frac{f}{\binom{K}{\Gamma}} \ \text{bits}.
\end{align}

To deliver $\{X_{\psi}\}_{\psi \in  \Psi_{\Gamma+1}}$, we will employ the last $K-\Gamma$ phases (phase $j = \Gamma+1, \dots,K$) of the MAT algorithm. 
Each phase $j$ delivers order-$j$ folded messages. We describe the content that is carried during each of these phases.

\emph{Phase $\Gamma+1$:} In this first phase of duration $T_{\Gamma+1}$, the information in $\{X_{\psi}\}_{\psi \in \Psi_{\Gamma+1}}$ is delivered by $\xv_{t}, \ t\in[0,T_{\Gamma+1}] $, which can also be rewritten in the form of a sequential transmission of shorter-duration $K$-length vectors $
\xv_{\psi} = [x_{\psi,1}, \dots, x_{\psi,K-\Gamma}, 0, \dots, 0]^{T}$ for different $\psi$, where each vector $\xv_{\psi}$ carries exclusively the information from each $X_{\psi}$, and where this information is uniformly split among the $K-\Gamma$ independent scalar entries $x_{\psi,i}, \ i=1,\dots,K-\Gamma$, each carrying $\frac{|X_{\psi}|}{(K-\Gamma)} = \frac{f}{\binom{K}{\Gamma}(K-\Gamma)} \ \text{bits}$ (cf.~\eqref{eq:XpsiSize}).
Hence, the duration of each $\xv_{\psi}$ is $\text{dur}(\xv_{\psi}) = \frac{|X_{\psi}|}{(K-\Gamma)f} = \frac{1}{\binom{K}{\Gamma}(K-\Gamma)}.$
Given that $|\mathcal{X}_\Psi|=\binom{K}{\Gamma+1}$, then
\begin{align} \label{eq:durationcal1}
T_{\Gamma+1} = \binom{K}{\Gamma+1} \text{dur}(\xv_{\psi}) =\frac{1}{\Gamma+1}.
\end{align}

After each transmission of $\xv_{\psi}$, each user $k\in[K]$ receives a linear combination $L_{\psi,k}$ of the transmitted $K-\Gamma$ symbols $x_{\psi,1}, x_{\psi,2}, \dots, x_{\psi,K-\Gamma}$.

Next an additional $K-\Gamma-1$ signals $L_{\psi,k'}, \ k'\in[K]\backslash \psi$ (linear combinations of $x_{\psi,1}, x_{\psi,2}, \dots, x_{\psi,K-\Gamma}$ as received --- up to noise level --- at each user $k'\in [K] \backslash \psi$) will be sent, which will help each user $k\in \psi$ to resolve $x_{\psi,1}, x_{\psi,2}, \dots, x_{\psi,K-\Gamma}$. This will be done in the next phase $j=\Gamma+2$.

\emph{Phase $\Gamma+2$:} The challenge now is for signals $$\xv_{c,t}, \ t \in (T_{\Gamma+1},T_{\Gamma+1}+T_{\Gamma+2}] $$ to convey all the messages of the form
$L_{\psi,k'}, \ \forall k'\in[K]\backslash \psi, \ \forall \psi\in \Psi_{\Gamma+1} $
to each receiver $k\in \psi$. Note that each of the above linear combinations, is now --- during this phase --- available (up to noise level) at the transmitter. Let
\begin{align}
\Psi_{\Gamma+2} = \{\psi\in [K] \ : \  |\psi|=\Gamma+2  \}
\end{align}
and consider for each $\psi\in \Psi_{\Gamma+2}$, a transmitted vector
\[\xv_\psi = [x_{\psi,1}, \dots, x_{\psi,K-\Gamma-1}, 0, \dots, 0]^{T}\]
which carries the contents of $\Gamma+1$ different linear combinations $f_i(\{L_{\psi\backslash\{k\},k}\}_{k\in \psi}), i=1,\dots,\Gamma+1$ of the $\Gamma+2$ elements $\{L_{\psi\backslash\{k\},k}\}_{\forall k\in \psi}$ created by the transmitter. The linear combination coefficients defining each linear-combination function $f_i$, are predetermined and known at each receiver. The transmission of $\{\xv_{\psi}\}_{\forall \psi \in \Psi_{\Gamma+2}}$ is sequential.

It is easy to see that there is a total of $(\Gamma+1)\binom{K}{\Gamma+2}$ symbols of the form $f_i (\{L_{\psi\backslash\{k\},k}\}_{k\in \psi}), i=1,\dots,\Gamma+1, \ \psi\in \Psi_{\Gamma+2}$, each of which can be considered as an order-$(\Gamma+2)$ signal intended for $\Gamma+2$ receivers in $\psi$. Using this, and following the same steps used in phase $\Gamma+1$, we calculate that
\begin{align} \label{eq:durationcal2}
T_{\Gamma+2} = \binom{K}{\Gamma+2}\text{dur}(\xv_{\psi}) = T_{\Gamma+1} \frac{\Gamma+1}{\Gamma+2}.
\end{align}

We now see that for each $\psi$, each receiver $k \in \psi$ recalls their own observation $L_{\psi \backslash \{k\}, k}$ from the previous phase, and removes it from all the linear combinations $\{ f_i (\{L_{\psi\backslash\{k\},k}\}_{\forall  k\in \psi})\}_{i=1,\dots,\Gamma+1}$, thus now being able to acquire the $\Gamma+1$ independent linear combinations $\{L_{\psi \backslash \{k'\}, k'}\}_{\forall  k' \in \psi \backslash \{k\}}$. It holds for each other user $k'\in \psi$.

After this phase, we use $L_{\psi,k}, \psi \in \Psi_{\Gamma+2}$ to denote the received signal at receiver $k$. Like before, each receiver $k, k\in \psi$ needs $K-\Gamma-2$ extra observations of $x_{\psi,1}, \dots, x_{\psi,K-\Gamma-1}$ which will be seen from $L_{\psi,k'},\forall k'\notin \psi$, which will come from order-$(\Gamma+3)$ messages that are created by the transmitter and which will be sent in the next phase.

\emph{Phase $j$ $(\Gamma+3 \leq j \leq K)$:}
Generalizing the described approach to any phase $j\in[\Gamma+3,\dots,K]$, we will use $\xv_{c,t}, \ t \in (\sum_{i=\Gamma+1}^{j-1}T_{i},\sum_{i=\Gamma+1}^{j}T_{i} ] $ to convey all the messages of the form
$L_{\psi,k'}, \ \forall k'\in[K]\backslash \psi, \ \forall \psi\in \Psi_{j-1} $
to each user $k\in \psi$. For each $\psi\in \Psi_{j} := \{\psi\in [K] \ : ~ |\psi|=j  \}$, each transmitted vector $\xv_\psi = [x_{\psi,1}, \dots, x_{\psi,K-j-1}, 0, \dots, 0]^{T}$
will carry the contents of $j-1$ different linear combinations $f_i(\{L_{\psi\backslash\{k\},k}\}_{k\in \psi}), i=1,\dots,j-1$ of the $j$ elements $\{L_{\psi\backslash\{k\},k}\}_{\forall k\in \psi}$ created by the transmitter. After the sequential transmission of $\{\xv_{\psi}\}_{\forall \psi \in \Psi_{j}}$, each receiver $k$ can obtain the $j-1$ independent linear combinations $\{L_{\psi \backslash \{k'\}, k'}\}_{\forall  k' \in \psi \backslash \{k\}}$. The same holds for each other user $k'\in \psi$.
As with the previous phases, we can see that
\begin{align} \label{eq:durationPhaseJ}
T_j = T_{\Gamma+1} \frac{\Gamma+1}{j}, \ j= \Gamma+3,\dots,K.
\end{align}
This process terminates with phase $j = K$, during which each $\xv_\psi = [x_{\psi,1}, 0 , 0, \dots, 0]^{T}$ carries a single scalar that is decoded easily by all. Based on this, backwards decoding will allow for users to retrieve $\{ X_{\psi}\}_{\psi \in \Psi_{\Gamma+1}}$. This is described below.

\subsection{Decoding}

Each receiver $k$ will backwards reconstruct the sets of overheard equations, which now take the form
\[\ba{c}
\{L_{\psi,k'}, \ \forall k'\in[K]\backslash \psi\}_{\forall \psi\in \Psi_{K}}  \\ \vdots \\ \downarrow \\ \{L_{\psi,k'}, \ \forall k'\in[K]\backslash \psi\}_{\forall \psi\in \Psi_{\Gamma+2}}  \ea
\]
until phase $\Gamma+2$, thus gaining enough observations to recover the original $K-\Gamma$ symbols $x_{\psi,1}, x_{\psi,2}, \dots, x_{\psi,K-\Gamma}$ that fully convey $X_{\psi}$, hence each user $k$ can reconstruct their own set $\{W_{R_k,\psi \backslash \{k\}}\}_{\psi \in  \Psi_{\Gamma+1}}$ by using $Z_{k}$, which in turn allows each user $k$ to reconstruct their requested $W_{R_k}$.

\subsection{Calculation of $T$}

To calculate $T$, combining \eqref{eq:durationcal1}, \eqref{eq:durationcal2} and \eqref{eq:durationPhaseJ} gives that $
T = \sum \limits_{j=\Gamma+1}^{K} T_j = T_{\Gamma+1}\sum \limits_{j=\Gamma+1}^{K} \frac{\Gamma+1}{j} \nonumber = \sum \limits_{j=\Gamma+1}^{K} \frac{1}{j}   = H_K-H_\Gamma $.

\section{Conclusions}
The work explored the interesting connections between retrospective transmission schemes which alleviate the effect of the delay in knowing the channel, and coded caching schemes which alleviate the effect of the delay in knowing the content destination. These connections are at the core of the coded caching paradigm, and their applicability can extend to different settings. For the MISO-BC setting, the optimal cache-aided DoF were identified within a multiplicative factor of 4. The result also implies that a very modest amount of caching can have a substantial impact on performance, as well as can go a long way toward removing the burden of acquiring timely CSIT.


\section*{Acknowledgment}
The work was supported by the ANR Jeunes Chercheurs project ECOLOGICAL-BITS-AND-FLOPS.

\section{Appendix - Proof of Lemma~\ref{lem:outer}\\ (Lower bound on $T^*$)}
Our aim is to lower bound the duration $T$, that guarantees the delivery of $K$ different files to $K$ users, via a MISO broadcast channel with delayed CSIT, and in the presence of $K$ caches, each of size $Mf$. Let $T_2$ be the duration needed to resolve the simpler setting where we want to serve $s\leq K $ different files to $s$ users, again each in the presence of their own caches. Naturally $T_2 \leq T$ since we ignore the interference from the remaining $K-s$ users (whose requests are ignored). Now let $T_3~(T_3 \leq T_2)$, be the duration needed to resolve the same problem, except that now all the $s$ caches are merged, and each of the $s$ users has access to all $s$ caches.
We choose to repeat this last experiment $\lfloor \frac{N}{s} \rfloor$ times, thus spanning a total duration of $T_3 \lfloor \frac{N}{s} \rfloor$. At this point, we transfer to the equivalent setting of the $s$-user MISO BC with delayed CSIT, and a side-information multicasting link to the receivers, of capacity $d_m$ (files per time slot). Under the assumption that in this latter setting, decoding happens at the end of communication, and once we set
\begin{align} \label{eq:DEFdm}
d_m T_3 \lfloor \frac{N}{s} \rfloor = sM\end{align}
(which guarantees that the side information from the side link, throughout the communication process, matches the maximum amount of information in the caches), we then have that \begin{align}
T_3 \lfloor \frac{N}{s}  \rfloor d'(d_m) \geq \lfloor \frac{N}{s} \rfloor s
\end{align}
where $d'(d_m)$ is any upper bound on the above $s$-user MISO BC channel with delayed CSIT and the side link. Using the bound 
\[d'(d_m) = \frac{s}{H_s}(1+d_m)\] from \cite{CYOG:14} and applying \eqref{eq:DEFdm}, we get
\[d'(d_m) = \frac{s}{H_s}(1+\frac{sM}{T_3 \lfloor \frac{N}{s} \rfloor })\] and thus we get
\begin{align}
T_3 \lfloor \frac{N}{s}  \rfloor \frac{s}{H_s}(1+\frac{sM}{T_3 \lfloor \frac{N}{s} \rfloor }) \geq \lfloor \frac{N}{s} \rfloor s
\end{align}
which means that 
\begin{align}
T_3 \geq H_s-\frac{sM}{ \lfloor \frac{N}{s} \rfloor}
\end{align}
which implies that the optimal $T^*$, for the original $s$-user problem, is bounded as
\begin{align}
T^*  \geq T_3 \geq H_s-\frac{sM}{ \lfloor \frac{N}{s} \rfloor}.
\end{align}
Maximization over all $s$, gives the desired result.

\section{Appendix - Bounding the gap to optimal\label{sec:gapCalculation}}
This section presents the proof that the gap $\frac{T(\gamma)}{T^*(\gamma)}$, between the achievable $T(\gamma)$ and the optimal $T^*(\gamma)$, is always upper bounded by 4, which also serves as the proof of identifying the optimal $T^*(\gamma)$ within a factor of 4.

First recall from Theorem~\ref{thm:MATandMN} that \[T(\gamma) = H_{K}-H_{K\gamma}\] and from Lemma~\ref{lem:outer} that \[T^*(\gamma) \geq \mathop {\text{max}}\limits_{s\in \{1, \dots, \lfloor \frac{N}{M} \rfloor\}} H_s-\frac{Ms}{\lfloor \frac{N}{s} \rfloor}. \]
We want to prove that
\begin{align}
\label{eq:seekBound1}
\frac{T(\gamma)}{T^*(\gamma)}< 4,  \ \forall K, \forall \Gamma = 1,2,\dots,K-1\end{align}
and the proof will be split into three cases: case 1 for $\gamma\in [\frac{1}{K}, \frac{1}{36}]$, case 2 for $\gamma\in [\frac{1}{36},\frac{1}{2}]$, and case~3 for $\gamma\in [\frac{1}{2},\frac{K-1}{K}]$. Recall that $\gamma$ is bounded as $\gamma\geq \frac{1}{K}$.

\subsubsection{Case 1 ($\gamma \leq \frac{1}{36}$)} First note that having $\gamma \leq \frac{1}{36}$ implies $K \geq 36$. To prove \eqref{eq:seekBound1}, we see that
\begin{align}
\frac{T}{T^*} &\leq \max_{\gamma\in[\frac{1}{K},\frac{1}{36}]\cap (\Z/K) }\frac{H_{K}-H_{K\gamma}}{\max\limits_{s\in[1,K]\cap \Z } H_s(1-\frac{Ms}{H_s \lfloor \frac{N}{s} \rfloor})} \label{eq:bound1a} \\
&\leq \max_{\gamma\in[\frac{1}{K},\frac{1}{36}]\cap (\Z/K) }\frac{H_{K}-H_{K\gamma}}{\max\limits_{s\in[6,\lfloor \sqrt{K} \rfloor]\cap \Z } (H_s-\frac{Ms}{\lfloor \frac{N}{s} \rfloor})} \label{eq:bound1a} \\
& \leq  \max_{\gamma\in[\frac{1}{K},\frac{1}{36}] }\frac{H_{K}-H_{K\gamma}}{\max\limits_{s\in[6,\lfloor \sqrt{K} \rfloor]\cap \Z } (H_s -\frac{Ms}{\lfloor \frac{N}{s} \rfloor})}
\label{eq:bound1b}\\
& \leq  \max_{\gamma\in[\frac{1}{K},\frac{1}{36}] } \frac{\log(\frac{1}{\gamma}) +\epsilon_{36}-\epsilon_\infty }{\max\limits_{s\in[6,\lfloor \sqrt{K} \rfloor]\cap \Z } \log(s) + \epsilon_\infty -\gamma s^2 \frac{7}{6}}\label{eq:bound1c}
 \end{align}
where \eqref{eq:bound1a} holds because $H_s-\frac{Ms}{ \lfloor \frac{N}{s} \rfloor}<0$ when $s>\lfloor \frac{1}{\gamma}\rfloor$ and because we reduced the maximizing region for $s$, where \eqref{eq:bound1b} holds because we increased the maximizing region for $\gamma$, and where~\eqref{eq:bound1c} holds because $\epsilon_K$ decreases with $K$, because $H_K-\log(K) \leq \epsilon_{36}$, $H_{K\gamma}-\log(K\gamma) > \epsilon_\infty$, $H_{s} > \log(s) +\epsilon_\infty$, and because $(\lfloor \frac{N}{s} \rfloor)/ \frac{N}{s} \geq \frac{6}{7}, \ s\leq \frac{N}{6}$ (recall that $s\leq \lfloor \sqrt{K} \rfloor \leq \frac{K}{6} \leq \frac{N}{6}$).
Continuing from~\eqref{eq:bound1c}, we have that
\begin{align}
\frac{T}{T^*} & \leq  \max_{s_c\in[6,\lfloor \sqrt{K} \rfloor]\cap \Z }   \max_{\gamma\in[\frac{1}{(s_c+1)^2},\frac{1}{s_c^2}] } \frac{\log(\frac{1}{\gamma}) +\epsilon_{36}-\epsilon_\infty }{\log (s_c) + \epsilon_\infty-\gamma s_c^{2} \frac{7}{6}}\label{eq:bound2a}
 \end{align}
 because \[\max\limits_{s\in[6,\lfloor \sqrt{K} \rfloor]\cap \Z } \log(s) + \epsilon_\infty -\gamma s^2 \frac{7}{6} \geq \log(s_c) + \epsilon_\infty-\gamma s_c^{2} \frac{7}{6}\] for any $\gamma$ and for any $s_c\in[6,\lfloor \sqrt{K} \rfloor]\cap \Z$. The split of the maximization $\max_{\gamma\in[\frac{1}{K},\frac{1}{36}] }$ into the double maximization $\max_{s_c\in[6,\lfloor \sqrt{K} \rfloor]\cap \Z }   \max_{\gamma\in[\frac{1}{(s_c+1)^2},\frac{1}{s_c^2}] }$ reflects the fact that we heuristically choose\footnote{Essentially we choose an $s$ that is approximately equal to $\lfloor \sqrt{\frac{1}{\gamma}} \rfloor$, and while this choice does not guarantee the exact maximizing $s$, it does manage to sufficiently raise the resulting lower bound.} $s = s_c\in \mathbb{Z}$ when $\gamma\in[\frac{1}{(s_c+1)^2},\frac{1}{s_c^2}]$.
 Now we perform a simple change of variables, introducing a real valued $s'$ ($s':=\sqrt{\frac{1}{\gamma}}$) such that $\gamma = \frac{1}{s'^2}$. Hence, a $\gamma$ range of $\gamma\in[\frac{1}{(s_c+1)^2},\frac{1}{s_c^2}]$, corresponds to an $s'$ range of $s'\in[s_c,s_c+1]$.
Hence we rewrite~\eqref{eq:bound2a} using this change of variables, to get
\begin{align}
\frac{T}{T^*} & \leq  \max_{s_c\in[6,\lfloor \sqrt{K} \rfloor]\cap \Z }   \max_{s'\in[s_c,s_c+1] } \frac{\log(s'^2) +\epsilon_{36}-\epsilon_\infty }{\log (s_c) + \epsilon_\infty -\frac{7}{6} \frac{s_c^2}{s'^2}} \\
& \leq  \max_{s_c\in[6,\lfloor \sqrt{K} \rfloor]\cap \Z }   \underbrace{\frac{\log(s_c+1)^2 +\epsilon_{36}-\epsilon_\infty }{\log (s_c) + \epsilon_\infty -\frac{7}{6}}}_{f(s_c)} \label{eq:bound3a} \\
& \leq   \frac{2*\log(7) +\epsilon_{36}-\epsilon_\infty }{\log (6) + \epsilon_\infty-\frac{7}{6}} \label{eq:bound4a} \\
& < 4
\end{align}
where \eqref{eq:bound3a} holds because $\frac{s_c^2}{s'^2} \leq 1 $, where \eqref{eq:bound4a} holds because $f(s_c)$ is decreasing in $s_c$.

\subsubsection{Case 2 ($\gamma \in[\frac{1}{36},\frac{1}{2}]$)}
In the maximization of the lower bound, we will now choose $s=1$.

For $K \geq 2$, we have
\begin{align}
\frac{T}{T^*} \leq \frac{\log(\frac{1}{\gamma})+\epsilon_2-\epsilon_\infty}{1-\gamma} =: f(\gamma)
\end{align}
because $H_K-\log(K) \leq \epsilon_{2}$, $H_{K\gamma}-\log (K\gamma) > \epsilon_\infty, \forall K\geq 2$.

For the above defined $f{(\gamma)}$, we calculate the derivative to take the form
\[\frac{d f{(\gamma)}}{d \gamma} = \frac{\overbrace{1-\gamma^{-1}-\log(\gamma)+ \epsilon_2-\epsilon_\infty}^{f'_N(\gamma)}}{\underbrace{(1-\gamma)^2}_{f'_D(\gamma)}}\] where $f'_N(\gamma),f'_D(\gamma)$ respectively denote the numerator and denominator of this derivative.
Since $f'_D(\gamma) >0, \forall \gamma < 1$, and since
\[\frac{d f'_N(\gamma)}{d \gamma} = \gamma^{-2}-\gamma^{-1} \geq 0, \forall \gamma \in [\frac{1}{36}, \frac{1}{2}].\] To prove this, we use the following lemma, which we prove in Section~\ref{sec:ratioOfDerivatives} below.

\vspace{3pt}
\begin{lemma} \label{lem:ratioOfDerivatives}
Let $g'_N (\gamma)$ and $g'_D (\gamma)$ respectively denote the numerator and the denominator of the derivative $\frac{d g(\gamma)}{d \gamma}$ of some function $g(\gamma)$. If in the range $\gamma \in [\gamma_1, \gamma_2]$, $g'_N (\gamma)$ increases in $\gamma$, and if $g'_D (\gamma)>0$, then
\begin{align}
\max_{\gamma \in [\gamma_1, \gamma_2]} g(\gamma) = \max \{g(\gamma=\gamma_1), g(\gamma=\gamma_2)\} \label{eq7}.
\end{align}
\end{lemma}

\vspace{3pt}

We now continue with the main proof, and apply Lemma~\ref{lem:ratioOfDerivatives}, to get
\begin{align}
\max_{\gamma \in [\frac{1}{36}, \frac{1}{2}]} f(\gamma)= \max\{f(\frac{1}{2}), f(\frac{1}{36})\} <4
\end{align}
which directly shows the desired $\frac{T}{T^*} < 4$ for $\frac{1}{36}\leq \gamma \leq \frac{1}{2}, K \geq 2$.

\subsubsection{Case 3 ($\gamma\in[\frac{1}{2},\frac{K-1}{K}]$)}
In the maximization of the lower bound, we will again choose $s=1$. Considering that now $\gamma$ takes the values $\gamma  = \frac{j}{K}, \ j \in [\frac{K}{2}, K-1]\cap \mathbb{Z}$, we have
\begin{align}
\frac{T}{T^*} &\leq \frac{H_K-H_{K\gamma}}{1-\gamma} = \frac{H_K-H_{(K-j)}}{j/K} \notag \\
                    &=\frac{1}{j}(\frac{K}{K-j+1}+ \frac{K}{K-j+2}+\cdots+1) \notag \\
                    &=\frac{1}{j}(1+\frac{j-1}{K-(j-1)}+1+\frac{j-2}{K-(j-2)}+\cdots+1) \notag \\
						        &= 1+ \frac{1}{j}(\frac{j-1}{K-(j-1)}+\frac{j-2}{K-(j-2)}+\cdots+\frac{1}{K-1}) \notag \\
										&< 2 \notag
\end{align}
because $j \leq \frac{K}{2}$.

This completes the proof for the entire case where $\Gamma = 1,2,\dots,K-1$.

\subsection{Proof of Lemma~\ref{lem:ratioOfDerivatives} \label{sec:ratioOfDerivatives}}
We first note that the condition $\frac{d g'_N (\gamma)}{d \gamma} \geq 0$ implies that $g'_N (\gamma)$ is increasing in $\gamma$. We also note that $g'_D (\gamma) \geq 0 , \gamma \in [\gamma_1, \gamma_2]$ where naturally $\gamma_1\leq  \gamma_2$. We consider the following three cases.

\paragraph{Case 1 ($g'_N (\gamma_1) \geq 0$)} If $g'_N (\gamma_1) \geq 0$ then $g'_N (\gamma) \geq 0$ for any $\gamma \in [\gamma_1, \gamma_2]$, which in turn means that $\frac{d g(\gamma)}{d \gamma} = \frac{g'_N (\gamma)}{g'_D (\gamma)} \geq 0, \gamma \in [\gamma_1, \gamma_2]$. This gives the desired
\[\mathop{\text{max}}\limits_{\gamma \in [\gamma_1, \gamma_2]} g(\gamma) = g(\gamma_2).\]

\paragraph{Case 2 ($g'_N (\gamma_1) <0 ~\& ~g'_N (\gamma_2) \leq 0$)} For any $\gamma \in [\gamma_1, \gamma_2]$, then if $g'_N (\gamma_1) <0 ~\& ~g'_N (\gamma_2) \leq 0$ then $g'_N (\gamma) \leq 0$, thus $\frac{d g(\gamma)}{d \gamma} \leq 0$, which gives the desired
\[\mathop{\text{max}}\limits_{\gamma \in [\gamma_1, \gamma_2]} g(\gamma) = g(\gamma_1).\]

\paragraph{Case 3 ($g'_N (\gamma_1) < 0 ~\& ~g'_N (\gamma_2) > 0$)} For any $\gamma \in [\gamma_1, \gamma_2]$, then if $g'_N (\gamma_1) < 0 ~\& ~g'_N (\gamma_2) > 0$, there exists a unique $\gamma = \gamma' \in [\gamma_1, \gamma_2]$ such that $g'_N (\gamma') = 0$. Hence $\frac{d g(\gamma)}{d \gamma} \leq 0, \forall \gamma \in [\gamma_1, \gamma']$ and $\frac{d g(\gamma)}{d \gamma} \geq 0, \forall \gamma \in [\gamma', \gamma_2]$. Consequently we have the desired
\[\mathop{\text{max}}\limits_{\gamma \in [\gamma_1, \gamma_2]} g(\gamma) = \mathop{\text{max}} \{g(\gamma_1), g(\gamma_2)\}. \]

Combining the above three cases, yields the derived $\mathop{\text{max}}\limits_{\gamma \in [\gamma_1, \gamma_2]} g(\gamma) = \mathop{\text{max}} \{g(\gamma_1), g(\gamma_2)\}$ which completes the proof.

\nocite{ZE:15}

\bibliographystyle{IEEEtran}
\bibliography{IEEEabrv,final_refs}

\begin{thebibliography}{10}
\providecommand{\url}[1]{#1}
\csname url@samestyle\endcsname
\providecommand{\newblock}{\relax}
\providecommand{\bibinfo}[2]{#2}
\providecommand{\BIBentrySTDinterwordspacing}{\spaceskip=0pt\relax}
\providecommand{\BIBentryALTinterwordstretchfactor}{4}
\providecommand{\BIBentryALTinterwordspacing}{\spaceskip=\fontdimen2\font plus
\BIBentryALTinterwordstretchfactor\fontdimen3\font minus
  \fontdimen4\font\relax}
\providecommand{\BIBforeignlanguage}[2]{{%
\expandafter\ifx\csname l@#1\endcsname\relax
\typeout{** WARNING: IEEEtran.bst: No hyphenation pattern has been}%
\typeout{** loaded for the language `#1'. Using the pattern for}%
\typeout{** the default language instead.}%
\else
\language=\csname l@#1\endcsname
\fi
#2}}
\providecommand{\BIBdecl}{\relax}
\BIBdecl

\bibitem{ZEarxiv:15}
\BIBentryALTinterwordspacing
J.~Zhang and P.~Elia, ``Fundamental limits of cache-aided wireless {BC:}
  interplay of coded-caching and {CSIT} feedback,'' \emph{CoRR}, vol.
  abs/1511.03961, 2015. [Online]. Available:
  \url{http://arxiv.org/abs/1511.03961}
\BIBentrySTDinterwordspacing

\bibitem{MN14}
M.~Maddah-Ali and U.~Niesen, ``Fundamental limits of caching,''
  \emph{Information Theory, IEEE Transactions on}, vol.~60, no.~5, pp.
  2856--2867, May 2014.

\bibitem{WLTL:15}
\BIBentryALTinterwordspacing
S.~Wang, W.~Li, X.~Tian, and H.~Liu, ``Fundamental limits of heterogenous
  cache,'' \emph{CoRR}, vol. abs/1504.01123, 2015. [Online]. Available:
  \url{http://arxiv.org/abs/1504.01123}
\BIBentrySTDinterwordspacing

\bibitem{MND13}
\BIBentryALTinterwordspacing
M.~A. Maddah{-}Ali and U.~Niesen, ``Decentralized caching attains order-optimal
  memory-rate tradeoff,'' \emph{CoRR}, vol. abs/1301.5848, 2013. [Online].
  Available: \url{http://arxiv.org/abs/1301.5848}
\BIBentrySTDinterwordspacing

\bibitem{JTLC:14}
\BIBentryALTinterwordspacing
M.~Ji, A.~M. Tulino, J.~Llorca, and G.~Caire, ``Order optimal coded delivery
  and caching: Multiple groupcast index coding,'' \emph{CoRR}, vol.
  abs/1402.4572, 2014. [Online]. Available:
  \url{http://arxiv.org/abs/1402.4572}
\BIBentrySTDinterwordspacing

\bibitem{MN:15isit}
M.~A. Maddah-Ali and U.~Niesen, ``Cache-aided interference channels,'' in
  \emph{Proceedings of the IEEE International Symposium on Information Theory
  (ISIT'2015)}, Hong-Kong, China, 2015.

\bibitem{TW:15}
\BIBentryALTinterwordspacing
R.~Timo and M.~A. Wigger, ``Joint cache-channel coding over erasure broadcast
  channels,'' \emph{CoRR}, vol. abs/1505.01016, 2015. [Online]. Available:
  \url{http://arxiv.org/abs/1505.01016}
\BIBentrySTDinterwordspacing

\bibitem{GKY:15}
\BIBentryALTinterwordspacing
A.~Ghorbel, M.~Kobayashi, and S.~Yang, ``Cache-enabled broadcast packet erasure
  channels with state feedback,'' \emph{CoRR}, vol. abs/1509.02074, 2015.
  [Online]. Available: \url{http://arxiv.org/abs/1509.02074}
\BIBentrySTDinterwordspacing

\bibitem{JTLC:15}
\BIBentryALTinterwordspacing
M.~Ji, A.~M. Tulino, J.~Llorca, and G.~Caire, ``Caching-aided coded
  multicasting with multiple random requests,'' \emph{CoRR}, vol.
  abs/1511.07542, 2015. [Online]. Available:
  \url{http://arxiv.org/abs/1511.07542}
\BIBentrySTDinterwordspacing

\bibitem{SJTLD:15}
K.~Shanmugam, M.~Ji, A.~Tulino, J.~Llorca, and A.~Dimakis, ``Finite length
  analysis of caching-aided coded multicasting,'' 2015, submitted to
  \emph{{IEEE} Trans. Inform. Theory - July 2015}.

\bibitem{ZFE:15}
J.~Zhang, F.~Engelmann, and P.~Elia, ``Coded caching for reducing
  {CSIT}-feedback in wireless communications,'' in \emph{Proc. Allerton Conf.
  Communication, Control and Computing}, Monticello, Illinois, USA, Sep. 2015.

\bibitem{SMK:15}
\BIBentryALTinterwordspacing
S.~P. Shariatpanahi, A.~S. Motahari, and B.~H. Khalaj, ``Multi-server coded
  caching,'' \emph{CoRR}, vol. abs/1503.00265, 2015. [Online]. Available:
  \url{http://arxiv.org/abs/1503.00265}
\BIBentrySTDinterwordspacing

\bibitem{HA:2015}
\BIBentryALTinterwordspacing
H.~Ghasemi and A.~Ramamoorthy, ``Improved lower bounds for coded caching,''
  \emph{CoRR}, vol. abs/1501.06003, 2015. [Online]. Available:
  \url{http://arxiv.org/abs/1501.06003}
\BIBentrySTDinterwordspacing

\bibitem{WLG:15}
\BIBentryALTinterwordspacing
C.~Wang, S.~H. Lim, and M.~Gastpar, ``Information-theoretic caching: Sequential
  coding for computing,'' \emph{CoRR}, vol. abs/1504.00553, 2015. [Online].
  Available: \url{http://arxiv.org/abs/1504.00553}
\BIBentrySTDinterwordspacing

\bibitem{APPV:15}
\BIBentryALTinterwordspacing
A.~N., N.~S. Prem, V.~M. Prabhakaran, and R.~Vaze, ``Critical database size for
  effective caching,'' \emph{CoRR}, vol. abs/1501.02549, 2015. [Online].
  Available: \url{http://arxiv.org/abs/1501.02549}
\BIBentrySTDinterwordspacing

\bibitem{HuangWDY015}
\BIBentryALTinterwordspacing
W.~Huang, S.~Wang, L.~Ding, F.~Yang, and W.~Zhang, ``The performance analysis
  of coded cache in wireless fading channel,'' \emph{CoRR}, vol.
  abs/1504.01452, 2015. [Online]. Available:
  \url{http://arxiv.org/abs/1504.01452}
\BIBentrySTDinterwordspacing

\bibitem{BBD:15}
\BIBentryALTinterwordspacing
E.~Bastug, M.~Bennis, and M.~Debbah, ``A transfer learning approach for
  cache-enabled wireless networks,'' \emph{CoRR}, vol. abs/1503.05448, 2015.
  [Online]. Available: \url{http://arxiv.org/abs/1503.05448}
\BIBentrySTDinterwordspacing

\bibitem{MCOFBJ:14}
\BIBentryALTinterwordspacing
A.~F. Molisch, G.~Caire, D.~Ott, J.~R. Foerster, D.~Bethanabhotla, and M.~Ji,
  ``Caching eliminates the wireless bottleneck in video-aware wireless
  networks,'' \emph{CoRR}, vol. abs/1405.5864, 2014. [Online]. Available:
  \url{http://arxiv.org/abs/1405.5864}
\BIBentrySTDinterwordspacing

\bibitem{HKD:14}
\BIBentryALTinterwordspacing
J.~Hachem, N.~Karamchandani, and S.~N. Diggavi, ``Coded caching for
  heterogeneous wireless networks with multi-level access,'' \emph{CoRR}, vol.
  abs/1404.6560, 2014. [Online]. Available:
  \url{http://arxiv.org/abs/1404.6560}
\BIBentrySTDinterwordspacing

\bibitem{HKS:15}
J.~Hachem, N.~Karamchandani, and S.~Diggavi, ``Effect of number of users in
  multi-level coded caching,'' in \emph{Proc. {IEEE} Int. Symp. Information
  Theory {(ISIT)}}, Hong-Kong, China, 2015.

\bibitem{DBAD:15}
M.~Deghel, E.~Bastug, M.~Assaad, and M.~Debbah, ``On the benefits of edge
  caching for {MIMO} interference alignment,'' in \emph{Signal Processing
  Advances in Wireless Communications (SPAWC), 2015 IEEE 16th International
  Workshop on}, June 2015, pp. 655--659.

\bibitem{MAT:11c}
M.~A. Maddah-Ali and D.~N.~C. Tse, ``Completely stale transmitter channel state
  information is still very useful,'' \emph{IEEE Trans. Inf. Theory}, vol.~58,
  no.~7, pp. 4418 -- 4431, Jul. 2012.

\bibitem{YKGY:12d}
S.~Yang, M.~Kobayashi, D.~Gesbert, and X.~Yi, ``Degrees of freedom of time
  correlated {MISO} broadcast channel with delayed {CSIT},'' \emph{IEEE Trans.
  Inf. Theory}, vol.~59, no.~1, pp. 315--328, Jan. 2013.

\bibitem{CE:13it}
J.~Chen and P.~Elia, ``Toward the performance versus feedback tradeoff for the
  two-user miso broadcast channel,'' \emph{IEEE Trans. Inf. Theory}, vol.~59,
  no.~12, pp. 8336--8356, Dec. 2013.

\bibitem{GJ:12o}
T.~Gou and S.~Jafar, ``Optimal use of current and outdated channel state
  information: {Degrees} of freedom of the {MISO} {BC} with mixed {CSIT},''
  \emph{IEEE Communications Letters}, vol.~16, no.~7, pp. 1084 -- 1087, Jul.
  2012.

\bibitem{CE:12d}
J.~Chen and P.~Elia, ``Degrees-of-freedom region of the {MISO} broadcast
  channel with general mixed-{CSIT},'' in \emph{Proc. Information Theory and
  Applications Workshop {(ITA)}}, Feb. 2013.

\bibitem{KYG:13}
P.~de~Kerret, X.~Yi, and D.~Gesbert, ``On the degrees of freedom of the
  {K}-user time correlated broadcast channel with delayed {CSIT},'' Jan. 2013,
  available on arXiv:1301.2138.

\bibitem{CYE:13isit}
J.~Chen, S.~Yang, and P.~Elia, ``On the fundamental feedback-vs-performance
  tradeoff over the {MISO}-{BC} with imperfect and delayed {CSIT},'' Jul. 2013,
  in \emph{ISIT13}, available on arXiv:1302.0806.

\bibitem{VV:09}
C.~Vaze and M.~Varanasi, ``{The degree-of-freedom regions of MIMO broadcast,
  interference, and cognitive radio channels with no CSIT},'' \emph{IEEE Trans.
  Inf. Theory}, vol.~58, no.~8, pp. 5254 -- 5374, Aug. 2012.

\bibitem{TJSP:12}
R.~Tandon, S.~A. Jafar, S.~Shamai, and H.~V. Poor, ``On the synergistic
  benefits of alternating {CSIT} for the {MISO} {BC},'' \emph{IEEE Trans. Inf.
  Theory}, vol.~59, no.~7, pp. 4106 -- 4128, Jul. 2013.

\bibitem{LH:12}
N.~Lee and R.~W. {Heath Jr.}, ``Not too delayed {CSIT} achieves the optimal
  degrees of freedom,'' in \emph{Proc. Allerton Conf. Communication, Control
  and Computing}, Oct. 2012.

\bibitem{HC:13}
C.~Hao and B.~Clerckx, ``Imperfect and unmatched {CSIT} is still useful for the
  frequency correlated {MISO} broadcast channel,'' in \emph{Proc. {IEEE} Int.
  Conf. Communications {(ICC)}}, Budapest, Hungary, Jun. 2013.

\bibitem{KPR:99}
M.~R. Korupolu, C.~G. Plaxton, and R.~Rajaraman, ``Placement algorithms for
  hierarchical cooperative caching,'' in \emph{Proc. ACM-SIAM SODA}, Jan. 1999,
  pp. 586--595.

\bibitem{BGW:10}
S.~Borst, V.~Gupta, and A.~Walid, ``Distributed caching algorithms for content
  distribution networks,'' in \emph{INFOCOM, 2010 Proceedings IEEE}, Mar. 2010,
  pp. 1--9.

\bibitem{CYOG:14}
\BIBentryALTinterwordspacing
J.~Chen, S.~Yang, A.~{\"{O}}zg{\"{u}}r, and A.~Goldsmith, ``Achieving full dof
  in heterogeneous parallel broadcast channels with outdated {CSIT},''
  \emph{CoRR}, vol. abs/1409.6808, 2014. [Online]. Available:
  \url{http://arxiv.org/abs/1409.6808}
\BIBentrySTDinterwordspacing

\bibitem{ZE:15}
J.~Zhang and P.~Elia, ``Fundamental limits of cache-aided wireless {BC}:
  {Interplay} of coded-caching and {CSIT} feedback,'' August 25 2015,
  \emph{EURECOM report No. RR-15-307}, available on:
  http://www.eurecom.fr/publication/4723.

\end{thebibliography}

\end{document}